\documentclass{aipproc}
\layoutstyle{6x9}
\usepackage{graphicx}
\usepackage{amsmath,amssymb}
\usepackage{epstopdf}
\begin{document}
\title{Robust Constraints on Dark Matter Annihilation into Gamma Rays and Charged Particles}

\author{Thomas D. Jacques} {address={School of Physics, The University of Melbourne, Victoria 3010, 
Australia}}

\date{August 14, 2009}

\begin{abstract}
Using gamma-ray data from observations of
the Milky Way, Andromeda (M31), and the cosmic
background, we calculate conservative upper limits on the dark matter
self-annihilation cross section to a number of final states,
over a wide range of dark matter masses. We first constrain annihilation to a pair of monoenergetic gamma rays, $\langle \sigma_A v \rangle_{\gamma \gamma}$, and show that in general our results are unchanged for a broader annihilation spectrum, if at least a few gamma rays are produced with energies within a factor of a few from the dark matter mass. We then place constraints on the self-annihilation cross section to an electron-positron pair, $\langle\sigma_A v \rangle_{e^+e^-}$, using gamma rays produced via internal bremsstrahlung radiative corrections. We also place constraints on annihilation into the other charged leptons.
We make conservative assumptions about the astrophysical inputs,
and demonstrate how our derived bounds would be strengthened if
stronger assumptions about these inputs are adopted.
\end{abstract}


% 95.35.+d Dark Matter
% 95.85.Pw Gamma-ray
% 98.70.Vc Cosmic background radiations
% 98.62.Gq Galactic halos

\maketitle

%%%%%%%%%%%%%%%%%%%%%%%%%%%%%%%
\section{Introduction}
%%%%%%%%%%%%%%%%%%%%%%%%%%%%%%%

For thermal relic Dark Matter (DM), the self-annihilation cross section must be $\langle \sigma_A v \rangle \sim 3 \times 10^{-26}$
cm$^3$ s$^{-1}$ to obtain the observed relic abundance.
There are a number of ways in which the self-annihilation rate can be enhanced e.g., \cite{Hisano:2004ds,Feng:2008ya,Cirelli:2008pk,ArkaniHamed:2008qn,  Pospelov:2008jd,Nelson:2008hj,Cholis:2008qq,Bai:2008jt,Fox:2008kb}, while non-thermal relic DM may allow larger values for  $\langle \sigma_A v \rangle$ \cite{Das:2006ht,KKT}.
Self-annihilation can lead to indirect detection of DM through observation of the annihilation products. We constrain the self-annihilation cross section by considering the gamma rays produced via DM annihilation to a number of final states. 

We first focus on annihilation to a pair of monenergetic gamma rays. This annihilation channel has a very clean signature, with $E_\gamma=m_\chi$, and future observation of such a line could provide clear evidence for the DM mass. 
We compare the calculated signal with gamma-ray observations of the Galactic Center, Andromeda (M31), and the cosmic background, and place upper limits on the partial self-annihilation cross section, $\langle \sigma_A v \rangle_{\gamma \gamma}$. Our limits cover a broad DM mass range, from 20 keV up to 10 TeV. We show our constraints are applicable even if DM annihilates to a broader spectrum of gamma rays, in general constraining the total annihilation cross section. With a conservative choice for the branching ratio,  $Br(\gamma\gamma)=10^{-4}$, we constrain the \emph{total} annihilation cross section, $\langle \sigma_A v \rangle_{\rm total}=\langle \sigma_A v \rangle_{\gamma \gamma}/Br(\gamma\gamma)$. 

Next we place constraints on the self-annihilation cross section to an electron-positron pair, $\langle\sigma_A v \rangle_{e^+e^-}$. These particles will inevitably be accompanied by gamma rays through a process known as \emph{internal bremsstrahlung} (IB) or \emph{final state radiation}. This is the emission of a gamma-ray from one of the final state charged particles, occurring at the Feynman diagram level. See Refs.~\cite{IB,BBB,Birkedal:2005ep,BergstromRC} for a detailed discussion. We use these gamma rays to constrain $\langle\sigma_A v \rangle_{e^+e^-}$ by again comparing the calculated annihilation signal with the observed gamma-ray flux. We place corresponding constraints on annihilation to the other charged leptons. 
The IB gamma-ray spectrum per annihilation does not depend on the tree level cross section, so our analysis remains DM model independent. Because emission occurs at the Feynman diagram level, the gamma-ray flux from IB remains independent of uncertain astrophysical phenomenon such as the interstellar magnetic field or radiation field.

The $e^+e^-$ final state is especially interesting in light of the recent PAMELA/ATIC/ Fermi results suggesting an excess in the observed positron flux. There are now numerous DM models that can explain these results, many with a large branching ratio to charged leptons, e.g. \cite{Cirelli:2008pk,ArkaniHamed:2008qn,Pospelov:2008jd,Nelson:2008hj,Cholis:2008qq,Bai:2008jt,Fox:2008kb}. Our limits on $\langle \sigma_A v \rangle_{e^+e^-}$ directly constrain many models attempting to explain the positron excess.

We show how our results depend on the astrophysical uncertainties in the DM density profile. We are conservative in our analysis methods and input choices, and show how our results would be strengthened by more optimistic choices.
We compare our constraints with pre-existing bounds on the total annihilation cross section. We find that our constraints on  $\langle\sigma_A v \rangle_{\rm total}$ are particularly strong for small DM masses, while our constraints on  $\langle\sigma_A v \rangle_{\gamma\gamma}$ are stronger than those on  $\langle\sigma_A v \rangle_{e^+e^-}$ by a factor of $\lesssim 10^2$, as expected from the $\alpha\simeq 1/137$ suppression of the radiative corrections.

%%%%%%%%%%%%%%%%%%%%%%%%%%%%%%%
\section{Annihilation to $\gamma\gamma$}
%%%%%%%%%%%%%%%%%%%%%%%%%%%%%%%

Our analysis and results are described in detail in Mack \emph{et al.} \cite{MJBBY}.
The flux of gamma rays from DM annihilation in our halo is given by
\begin{equation}
\frac{d\Phi_{\gamma}}{dE} =
\frac{\langle \sigma_A v\rangle}{2}
\frac{{\cal J}_{\Delta \Omega}}{{\rm J_0}}
\frac{ 1 }{4\pi m_\chi^2}
\frac{dN_{\gamma}}{dE}\,,
\label{haloflux}
\end{equation}
where $dN_\gamma/dE$ is the gamma-ray spectrum per annihilation, $\Delta\Omega$ is the field of view in steradians, and 

\begin{equation}
{\cal J}_{\Delta \Omega} = \frac{2\pi }{\Delta \Omega} {\rm J_0}\int_0^{\psi} 
\int^{\ell_{max}}_0 \rho^2\left(\sqrt{R_{\textrm{sc}}^2 -
2\ell R_{\textrm{sc}}\cos{\psi} +\ell^2}\right)d\ell \sin{\psi }d\psi\,,
\end{equation}
is proportional to the dark matter density squared, integrated over the line of sight. %Our goal is to place robust upper limits on $\langle \sigma_A v\rangle$, and so we adopt the conservative Kravtsov DM density profile for the Milky Way \cite{}. We also show constraints using the more commonly adopted NFW profile \cite{}. Wherever possible we use flux data with a wide field of view, minimizing uncertainties in ${\cal J}_{\Delta \Omega}$ due to differences between density profiles.
 For annihilation to a gamma-ray pair, the spectrum per annihilation is simply $dN_{\gamma}/dE = 2\delta(m_\chi - E)$. 

We collect gamma-ray flux data from a variety of sources, covering a large range of energies. We compare the calculated annihilation flux with the observed flux, integrating each over an energy bin of equal size, and find an upper limit on $\langle \sigma_A v\rangle$ by demanding that the annihilation flux be smaller than the entire observed flux. This is extremely conservative, as in reality, the DM annihilation flux would only be a small fraction of the observed flux. % It is also robust, as the annihilation flux cannot be larger than the observed flux.

We focus on the Galactic Center flux, using data from COMPTEL and EGRET aboard the CGRO \cite{COMPTELweb,EGRETweb}, INTEGRAL \cite{INTweb},  and H.E.S.S. \cite{HESSweb}. We also look at gamma rays arising from annihilation in the M31 (Andromeda) galaxy, and from cosmic annihilation, using the methods described in Mack \emph{et al.} \cite{MJBBY}. 
We use M31 data from EGRET, CELESTE \cite{CELESTEweb} and HEGRA \cite{HEGRAweb}, and diffuse cosmic flux data from INTEGRAL, COMPTEL, EGRET and the Solar Maximum Mission \cite{SMMweb}. 

\begin{figure}
\includegraphics[width=3.25in, clip=true]{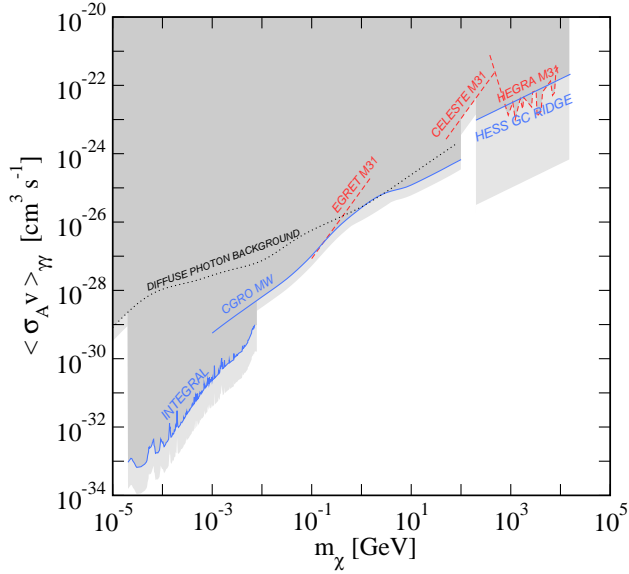}
\caption{
  The limits on the partial cross section, $\langle \sigma_A v
  \rangle_{\gamma\gamma}$, derived from the various gamma-ray data.
  Our overall limit is shown as the dark shaded exclusion region.  For
  comparison, the light-shaded region shows the corresponding limits
  for the NFW (rather than the Kravtsov) profile.}
\label{fig:sigmagg-fig2}
\end{figure}

We place constraints on  $\langle \sigma_A v\rangle_{\gamma\gamma}=\langle \sigma_A v\rangle_{\rm total} Br(\gamma\gamma)$, for both the conservative Kravtsov \cite{Kravtsov} and the commonly-adopted NFW DM density profile \cite{NFW}, over an energy range spanning 20 keV to 10 TeV. These results are shown in Fig.~\ref{fig:sigmagg-fig2}. 

%%%%%%%%%%%%%%%%%%%%%%%%%%%%%%%%
%%%%%%%%%%%%%%%%%%%%%%%%%%%%%%%%

\begin{figure}
\includegraphics[width=3.25in, clip=true]{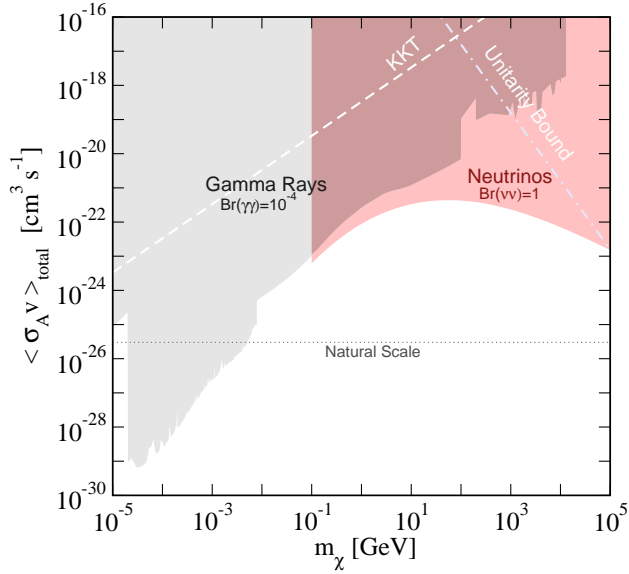}
\caption{
  The gamma-ray and neutrino limits on the total annihilation
  cross section, selecting $Br(\gamma\gamma) = 10^{-4}$ as a
  conservative value.  The unitarity and KKT bounds are also shown.
  The overall bound on the total cross section at a given mass is
  determined by the strongest of the various upper limits.}
  \label{fig:sigmagg-fig3}
\end{figure}

To find a constraint on the total cross section, we need the branching ratio to this final state. Typical branching ratios are $\sim 10^{-3}$.
In Fig.~\ref{fig:sigmagg-fig3} we show our upper bound on $\langle \sigma_A v\rangle_{\rm total}$ using a conservative choice of  $Br(\gamma\gamma)=10^{-4}$. In this plot we use the conservative Kravtsov density profile. For comparison, we show other limits on $\langle \sigma_A v\rangle$, including a bound based on observations of the Neutrino flux, from Y\"{u}ksel \emph{et al.} \cite{YHBA}, and the standard cross section for thermal relic DM. The KKT and Unitarity bounds are described in Ref.~\cite{MJBBY}.

%%%%%%%%%%%%%%%%%%%%%%%%%%%%%%%%
%%%%%%%%%%%%%%%%%%%%%%%%%%%%%%%%

Most of these experiments had modest energy resolution.  To
be conservative, we assume an analysis bin with a logarithmic
energy width of 0.4 in $\log_{10} E$ (i.e., $\Delta(\ln E) \sim 1$)
for the Galactic and cosmic diffuse analyses; this is at least as wide as the energy bins
reported by the experiments. In effect, our
results are what one would obtain for an annihilation gamma-ray
spectrum as  wide as 0.4 in $\log_{10} E$.  The exception is the
INTEGRAL line search.

%%%%%%%%%%%%%%%%%%%%%%%%%%%%%%%
\section{Annihilation to $e^+e^-$}
%%%%%%%%%%%%%%%%%%%%%%%%%%%%%%%

\begin{figure}
\includegraphics[width=3.0in, clip=true]{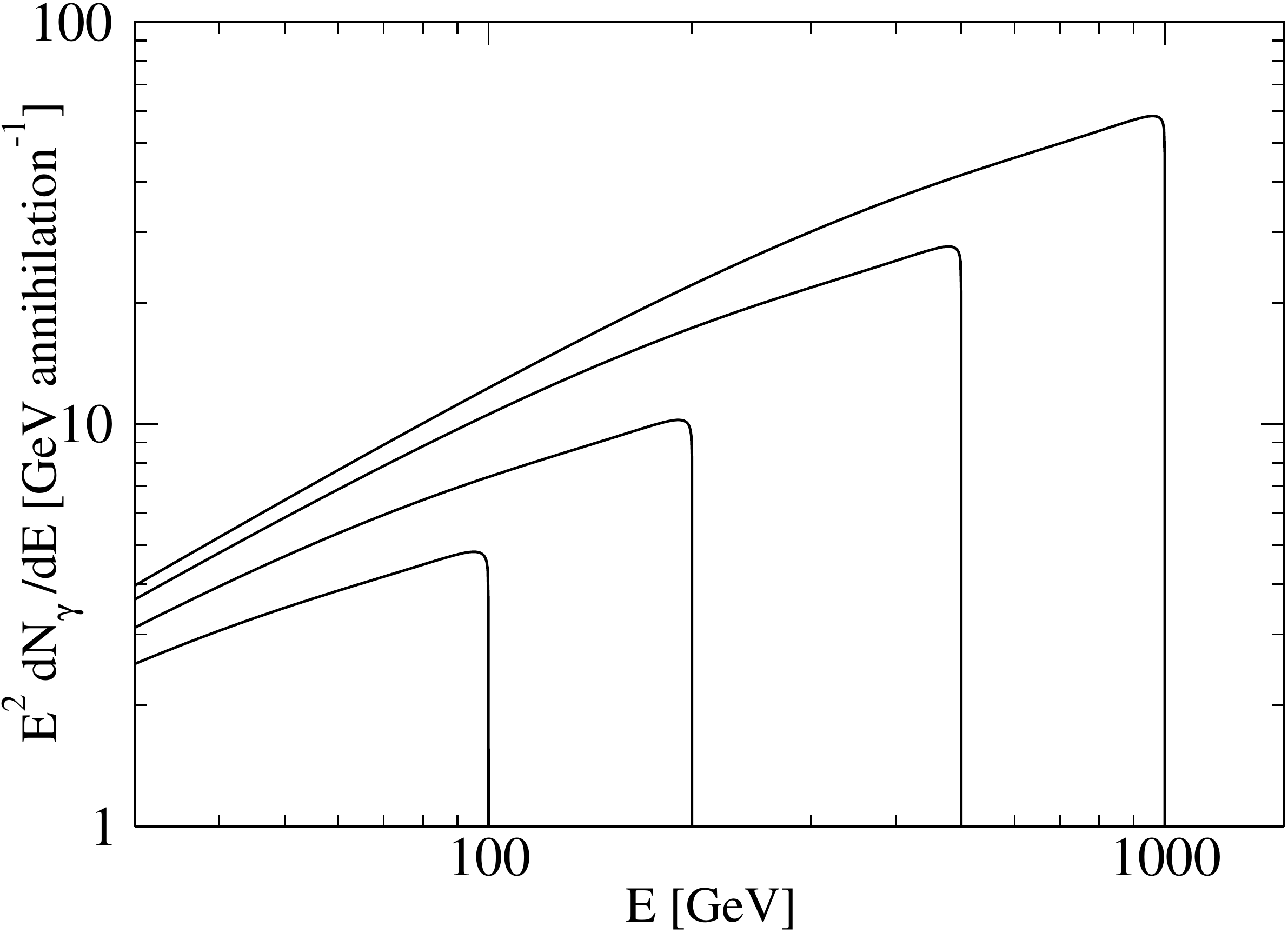}
\caption{ Internal bremsstrahlung gamma-ray spectra per
  $\chi\chi\rightarrow e^+e^-$ annihilation, for $m_\chi=$ 100 GeV,
  200 GeV, 500 GeV, 1000 GeV.}
\label{spectra}
\end{figure}

For annihilation to an $e^+e^-$ pair, we follow a similar method to that outlined above. We constrain $\langle \sigma_A v \rangle_{e^+e^-}$ by demanding that the calculated gamma-ray signal from IB emission not exceed the entire observed gamma-ray flux. Our analysis and results are described in detail in Bell and Jacques \cite{IB}. The IB gamma-ray spectrum per annihilation is
\begin{equation}
\frac{dN_\gamma}{dE} = 
\frac{1}{\sigma_{\rm tot}}\frac{d\sigma_{\rm IB}}{dE_\gamma}.
\label{Ngamma}
\end{equation}
where
\begin{equation}
\frac{d\sigma_{\rm IB}}{dE_\gamma} = \sigma_{\rm tot} \times 
\frac{\alpha}{E_\gamma \pi} 
\bigg[ \ln\bigg(\frac{4m_\chi(m_\chi-E_\gamma)}{m_e^2} \bigg) -1 \bigg]\bigg[1+\bigg( \frac{4m_\chi(m_\chi-E_\gamma)}{4m_\chi^2}\bigg)^2\bigg],
\label{sigbrem}
\end{equation}
and $\sigma_{\rm tot}$ is the
tree-level cross section for annihilation to $e^+e^-$.  Note that
$\sigma_{\rm tot}$ factors out from the IB cross-section.  This
important feature implies that the IB spectrum is independent of the
unknown physics which mediates the lowest order annihilation process.
Fig.~\ref{spectra} shows $E^2 dN_\gamma/dE$ for IB emission for a number of DM masses.

%%%%%%%%%%%%%%%%%%%%%%%%%%%%%%%%%%%%%%%%%%%%%%%%%%%
%%%%%%%%%%%%%%%%%%%%%%%%%%%%%%%%%%%%%%%%%%%%%%%%%%%

We use Galactic gamma-ray data from COMPTEL, EGRET and H.E.S.S. We also use data from CELESTE observations of M31 to cover the gap between the highest energy EGRET data and the lowest H.E.S.S. data points. 
The strongest bounds are obtained by choosing an analysis bin where the ratio of the calculated annihilation signal to the observed background is largest. Fig.~\ref{spectra} shows that $dN_\gamma/dE$ falls with energy. However, the observed gamma-ray flux falls with energy at a steeper rate, so that the strongest results are obtained by choosing an analysis bin with the DM mass as the upper energy limit.
%
%%%%%%%%%%%%%%%%%%%%%%%%%%%%%%%%%%%%%%%%%%%%%%%%%%%
  \begin{figure}
\includegraphics[width=3.5in, clip=true]{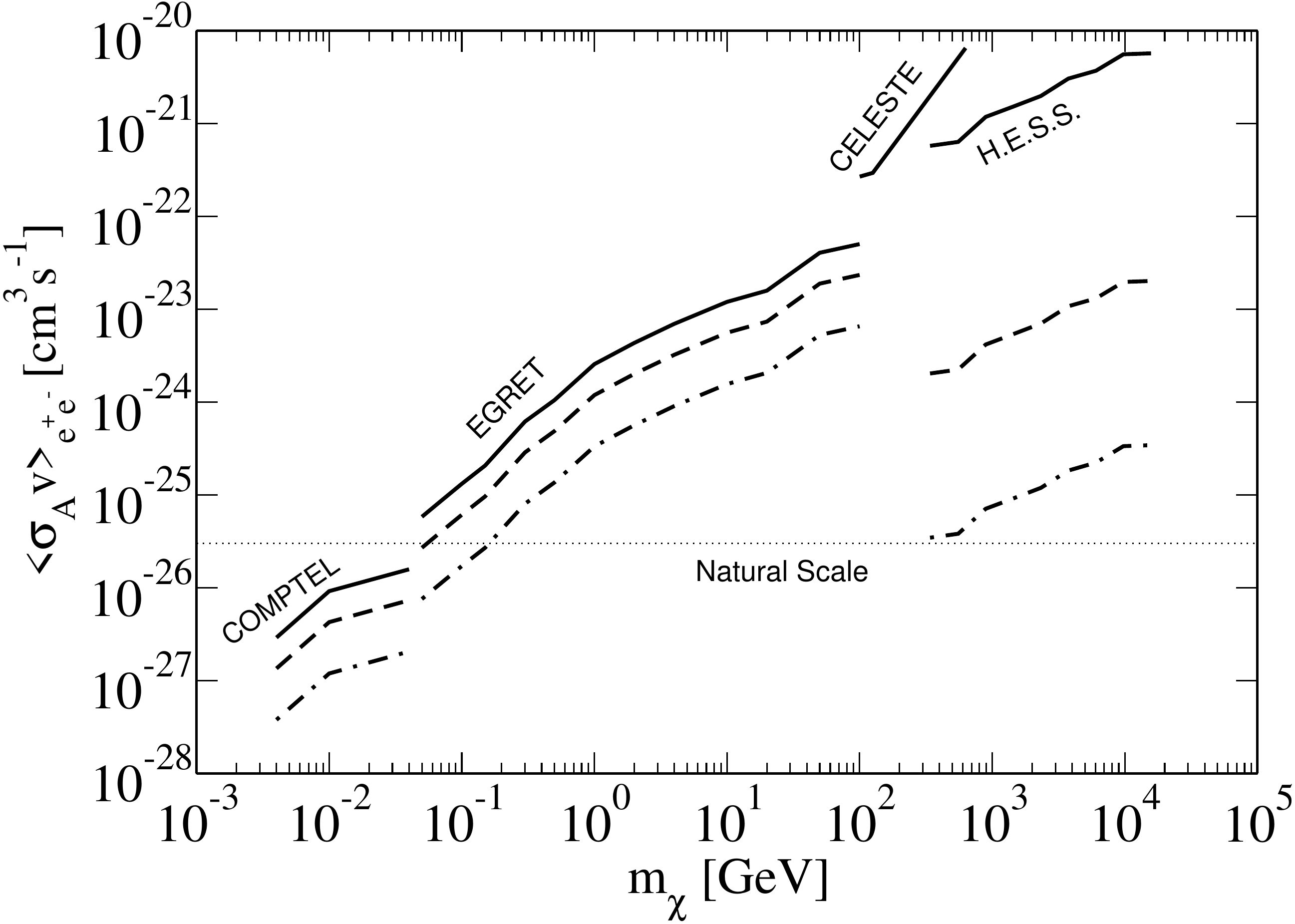}
\caption{Upper limit on $\langle \sigma v \rangle_{e^+e^-}$ as a
  function of DM mass for the Kravtsov (solid), NFW (dashed) and
  Moore (dot-dashed) profiles.}
\label{results1}
\end{figure}
%%%%%%%%%%%%%%%%%%%%%%%%%%%%%%%%%%%%%%%%%%%%%%%%%%%
In Fig.~\ref{results1} we show the upper limits on $\langle \sigma_A v
\rangle_{e^+e^-}$ as a function of DM mass, using the observational
data described above.  We give the Galactic Center results for the
Kravtsov and NFW profiles mentioned earlier, as well as the steep Moore profile \cite{Moore}. We show the CELESTE constraint
using only the Kravtsov profile.

\begin{figure}
\includegraphics[width=3.5in, clip=true]{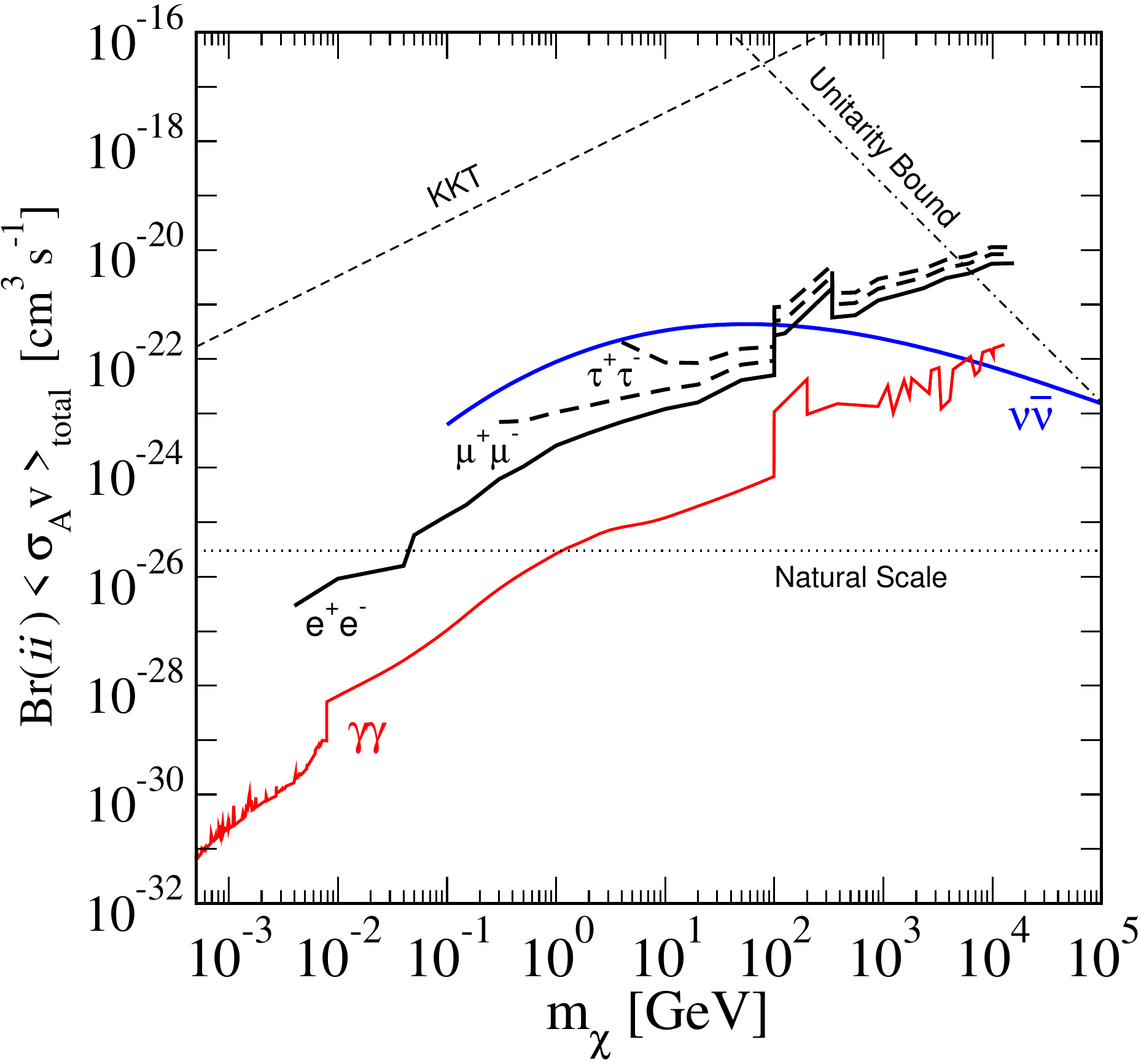}
\caption{Upper limits on $Br(ii)\times\langle \sigma v
  \rangle_{total}$ for various final states $ii=e^+e^-$ (solid black;
  labelled), $\mu^+\mu^-$ (thick dashed; labelled), $\tau^+\tau^-$ (thick
  dashed; labelled), $\gamma\gamma$ (solid; labelled), and $\bar\nu\nu$
  (solid; labelled), using the conservative Kravtsov profile. Also
  shown are the KKT (thin dashed) and unitarity (thin dot-dashed)
  limits on the total cross section described in Refs.~\cite{MJBBY,IB}, and the
  cross section for thermal relic DM (natural scale).  The
  $\bar\nu \nu$ limits are taken from
Ref.~\cite{YHBA}.}
\label{results2}
\end{figure}

In Fig.~\ref{results2} we show the upper bounds on the annihilation
cross sections into $e^+e^-$, $\mu^+\mu^-$, and $\tau^+\tau^-$, based
upon the IB emission from each final state (all use the conservative
Kravtsov profile). For annihilation to $\tau^+\tau^-$, we do not take into account gamma rays arising from the hadronic decay modes. Including this would strengthen our bounds.
For comparison, we also show limits on $\langle \sigma_A v \rangle_{\gamma\gamma}$ discussed in the previous section, as well as some earlier bounds on $\langle \sigma_A v\rangle$. For full details see Ref.~\cite{IB}.
%
%%%%%%%%%%%%%%%%%%%%%%%%%%%%%%%%%%%%%%%%%%%%%%%%%%%
%%%%%%%%%%%%%%%%%%%%%%%%%%%%%%%%%%%%%%%%%%%%%%%%%%%

%%%%%%%%%%%%%%%%%%%%%%%%%%%%%%%
\section{Discussion}
%%%%%%%%%%%%%%%%%%%%%%%%%%%%%%%

While the Kravtsov, NFW and Moore profiles
diverge towards the center of the Galaxy, they are similar at large
radii. As the INTEGRAL, EGRET and COMPTEL observations encompass relatively
large angular scales, the density profile changes have a modest
effect.  On the other hand, the H.E.S.S. constraints correspond to a much
smaller angular region toward the Galactic Center, and vary by orders
of magnitude depending on the profile adopted. (See Ref.~\cite{YHBA} for a full discussion of the differences between
the profiles for different angular regions.)
%
%To be conservative, we do not consider the possibility that DM
%annihilation rates are enhanced due to substructure in the halo, e.g.,
%Refs.~\cite{Diemand:2006ik,Strigari:2007at,Bi:2005im,Hooper:2007be},
%or mini-spikes around intermediate-mass black
%holes~\cite{Bertone:2005xz, Horiuchi:2006de}; such enhanced
%annihilation signals would result in stronger upper bounds on the
%cross section.

If DM annihilates to $e^\pm$, photons will be produced not only by IB,
but also by energy loss processes including inverse Compton scattering
and synchrotron radiation. However, the intensity of these signals depend on uncertain astrophysical parameters, such as magnetic field strength,
radiation field intensities, and electron diffusion scales.  By
contrast, IB is free of these astrophysical uncertainties, and has a
fixed spectrum and normalization.
Another key difference is the energy of the photons.  Synchrotron radiation
produces generally low energy photons, while IB provides some hard
gamma rays near the endpoint.  Since the background flux falls off
with energy, these hard gamma rays are extremely useful.
The sharp edge in the IB spectrum at $E=m_\chi$ can be used to
diagnose the DM mass; this is not possible with synchrotron radiation.

Refs.~\cite{Cirelli:2008pk,ArkaniHamed:2008qn,Pospelov:2008jd,Nelson:2008hj,Cholis:2008qq,Bai:2008jt,Fox:2008kb} and others
have recently proposed models in which DM annihilates directly to
charged leptons, with cross sections well above that expected for a
thermal relic.
This may account for anomalies in cosmic ray spectra from PAMELA,
ATIC and Fermi, and microwave signals from
WMAP, all of which seem to require more electrons and positrons than
can be explained otherwise.  Our bounds on $\langle \sigma_A v
\rangle_{l^+l^-}$ will directly constrain the allowed parameter space
for these types of DM models.

%%%%%%%%%%%%%%%%%%%%%%%%%%%%%%%%%%%%%%%%%%%%
%%%%%%%%%%%%%%%%%%%%%%%%%%%%%%%%%%%%%%%%%%%%

\medskip

{\bf Acknowledgments:}
This article is based on Ref.~\cite{MJBBY}, done in collaboration with Greg Mack, John Beacom, Nicole Bell, Hasan Yuksel, and Ref.~\cite{IB}, done in collaboration with Nicole Bell.
The author was supported by the Commonwealth of Australia.

%%%%%%%%%%%%%%%%%%%%%%%%%%%%%%%%%%%%%%%%%%%%
%%%%%%%%%%%%%%%%%%%%%%%%%%%%%%%%%%%%%%%%%%%%

\end{document}